\documentclass[12pt]{iopart}
\usepackage{iopams}  
\usepackage{graphicx}

\begin{document}

\title{Statistical Model Predictions for Pb-Pb Collisions at LHC}

\author{I~Kraus$^1$, J~Cleymans$^2$, H Oeschler$^1$, K~Redlich$^3$ and S~Wheaton$^{1,2}$}

\address{$^1$~ Institut f\"ur Kernphysik, Darmstadt University of Technology, D-64289 Darmstadt, Germany\\
$^2$~ UCT-CERN Research Centre and Department  of  Physics, University of Cape Town, Rondebosch 7701, South Africa\\
$^3$~ Institute of Theoretical Physics, University of Wroc\l aw, Pl-45204 Wroc\l aw, Poland}

\ead{Ingrid.Kraus@cern.ch}

\begin{abstract}
The systematics of Statistical Model parameters extracted from heavy-ion collisions at lower energies are exploited to extrapolate in the LHC regime. Predictions of various particle ratios are presented and particle production in central Pb-Pb collisions at LHC is discussed in the context of the Statistical Model. 
The sensitivity of several ratios on the temperature and the baryon chemical potential is studied in detail, and some of them, which are particularly appropriate to determine the chemical freeze-out point experimentally, are indicated. 
The impact of feed-down contributions from resonances, especially to light hadrons, is illustrated.
\end{abstract}

\maketitle

\section{Introduction}
Particle production in heavy-ion collision has been measured over a broad range of incident energies from SIS up to RHIC~\cite{sqm2004}.
Systematic studies of the thermal properties of the final state show that the particle yields are consistent with the assumption
that hadrons originate from a thermal source of a given temperature and baryon chemical potential~\cite{review}. 
The thermal parameters, quantified by comparing calculated particle yields and ratios to experimental data, feature a common
energy dependence, which can be characterized e.~g. by a constant average energy per hadron~\cite{param}. This parametrisation allows
for extrapolations in the LHC regime and for prediction for particle ratios presented in this paper.

\section{Assumptions and predictions}

The Statistical Model and its application to heavy-ion collisions has been summarized recently~\cite{review}.
In view of the large number of produced particles expected in Pb-Pb reactions at LHC, the grand canonical formulation is used. 
In this study, no particle-specific excluded-volume corrections were applied.
The quantum numbers of conserved charges, including strangeness, are controlled by the corresponding potentials. 
Moreover, the chemical potentials are constrained by the initial isospin asymmetry and strangeness neutrality of the incoming nuclei. Thus, any particle ratio is uniquely determined by only two parameters, the temperature $T$ and the baryon chemical potential $\mu_B$.
\\
The parameterized energy dependence of $T$ and $\mu_B$, as it was extracted recently from the Statistical Model analysis of hadron multiplicities~\cite{param}, allows to extrapolate $\mu_B$ to the LHC energy, resulting in $\mu_B(\sqrt{s_{NN}} = 5.5 \rm TeV) = 1~MeV$.
The chemical freeze-out temperature of $T$ = 170 MeV was chosen, which is consistent with the values derived at top SPS and at RHIC energies. To account for possible uncertainties in the extrapolation in the chemical freeze-out parameters, a possible shift of $T$ by 5 MeV and a variation of $\mu_B$ between 0 and 5 MeV is considered.
\\
\begin{table}
\begin{center}
\caption{\label{Table} Particle ratios in central Pb-Pb collisions at freeze-out conditions expected at the
					LHC: $T$ = (170$\pm$5) MeV  and $\mu_B$~=~${\rm 1^{+4}_{-1}}$ MeV. 
					The given errors correspond to the variation in the thermal parameters.
					Additional, systematic uncertainies in the ratios of the right column arise from
					unknown decay modes. They are smaller than 1\% in general, but reach 3\% in 
					the $\Xi^- / \Lambda$ ratio and 7\% in the  p$ / \pi^-$ and the $\Lambda / $p  
					ratios.}
\begin{tabular}{cc|cc}
\hline
\multicolumn{2}{c|}{ $\bar{h} / h$ Ratio } & \multicolumn{2}{c}{ mixed Ratio } \\ 
\hline
 $ \pi^+ / \pi^- ~ $ & ~ $ 0.9998^{+ 0.0002}_{- 0.0010}$ ~ & ~ $ \rm K^+ / \pi^+ ~ $ & ~ $ 0.180^{+ 0.001}_{- 0.001} $~  \\
 $ \rm K^+ / \rm K^- ~ $ & ~ $ 1.002^{+ 0.008}_{- 0.002}$ ~ & ~ $ \rm K^- / \pi^-~ $ & ~ $ 0.179^{+ 0.001}_{- 0.001} $~  \\
 $ \rm \bar{p} / \rm p	~ $ & ~ $ 0.989^{+ 0.011}_{- 0.045}$ ~ & ~ $ \rm p / \pi^-	~ $ & ~ $ 0.091^{+ 0.009}_{- 0.007} $~  \\
 $ \bar{\Lambda} / {\Lambda}~ $ & ~ $ 0.992^{+ 0.009}_{- 0.036}$ ~ &  ~ $ \Lambda / \rm p ~ $ & ~ $ 0.473^{+ 0.004}_{- 0.006} $~  \\
 $ \bar{\Xi}^+ / {\Xi}^-	~ $ & ~ $ 0.994^{+ 0.006}_{- 0.026}$ ~ &  ~ $ \Xi^- / \Lambda	~ $ & ~ $ 0.160^{+ 0.002}_{- 0.003} $~  \\
 $ \bar{\Omega}^+ / {\Omega}^-	~ $& ~ $ 0.997^{+ 0.003}_{- 0.015}$ ~ & ~ $ \Omega^- / \Xi^-~ $ & ~ $ 0.186^{+ 0.008}_{- 0.009} $~  \\
%
\hline
\end{tabular}
\end{center}
\end{table} 
Predictions for different particle ratios, calculated with the {\sc Thermus} package~\cite{thermus}, are presented in Table~\ref{Table}.
One of the expected features of particle production at LHC is that there should be a rather negligible difference between the yield of particles and their antiparticles. This is a direct consequence of the low net-baryon and net-charge density expected in the collision fireball at LHC at chemical decoupling.
On the other hand, the ratios of particles with different masses and different quantum numbers are predominantly controlled by the freeze-out temperature and the particle masses. The values of such yield ratios vary strongly with the particle mass difference, as seen in the right panel of Table \ref{Table}, see \cite{prc} for a detailed discussion. 

\section{Sensitivity to freeze-out parameters} 

\begin{figure}[htb]
\begin{minipage}[b]{0.53\linewidth}
\centering
\includegraphics*[width=\linewidth]{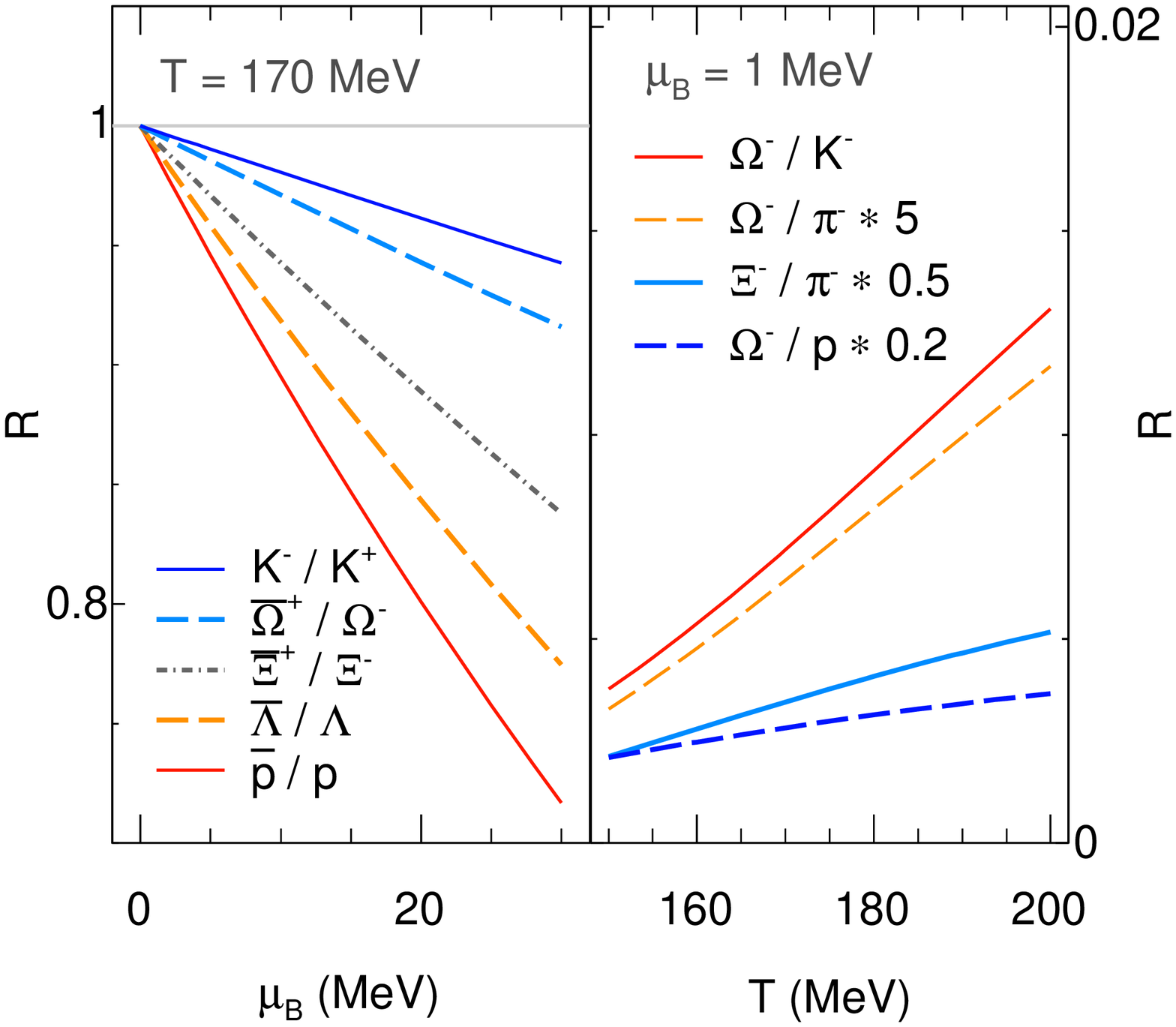}
\end{minipage}\hfill
\begin{minipage}[b]{0.47\linewidth}
\centering
\includegraphics[width=\linewidth]{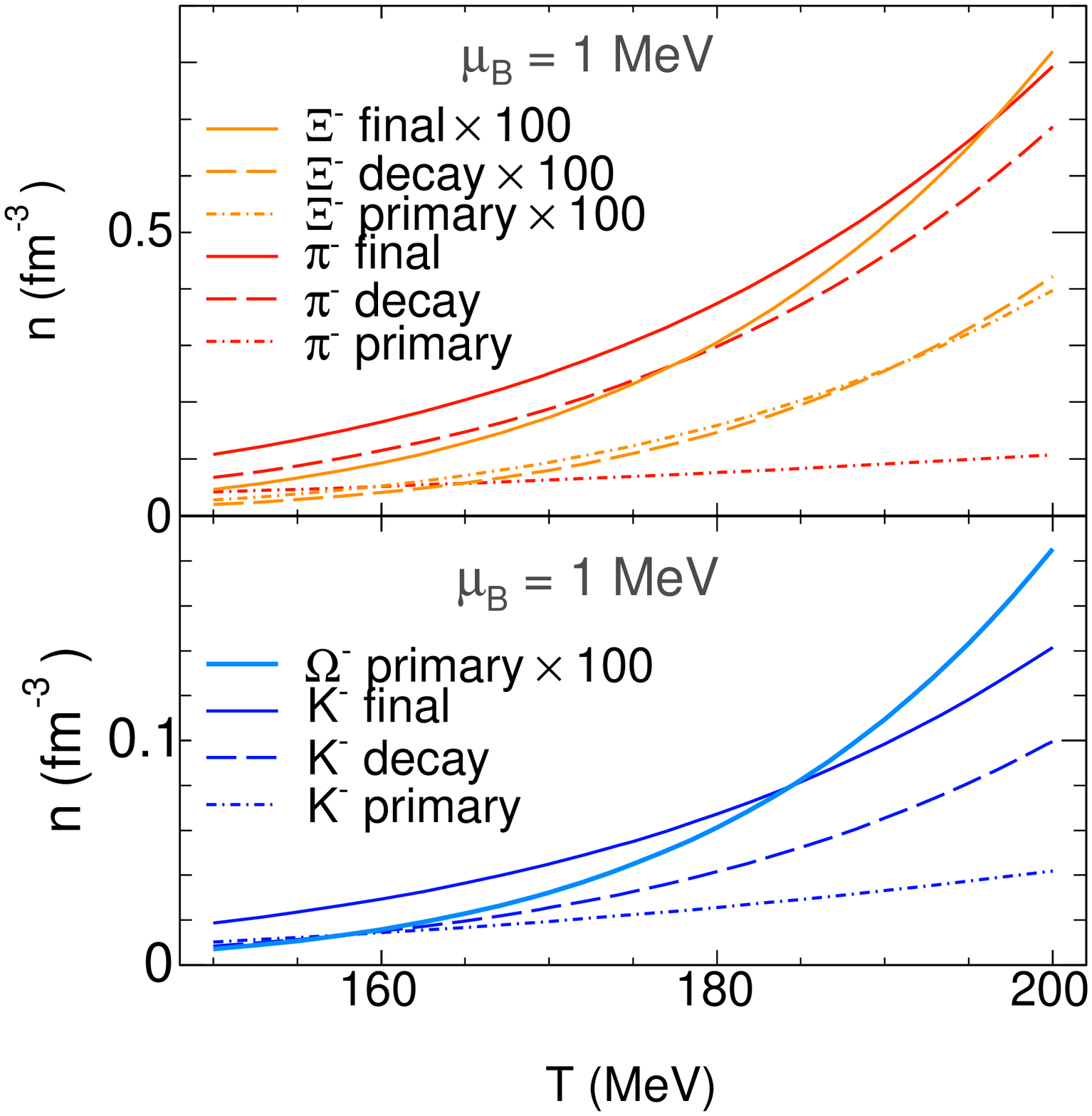}
\end{minipage}
\caption{\label{fig}Antiparticle/particle ratios $R$ as a function of $\mu_B$ for $T$ = 170~MeV (a),
		the horizontal line at 1 is meant to guide the eye.
		Particle ratios $R$ involving hyperons as a function of  $T$ for $\mu_B$ = 1 MeV (b).
Particle densities of $\pi^-$ and $\Xi^-$ (c) and of K$^-$ and $\Omega^-$ (d) hadrons as a function of $T$ for $\mu_B$ = 1 MeV. Different sources are indicated.} 
\end{figure}

The dependence of different particle ratios on the values of the thermal parameters is studied. The main objective is to identify observables  that could serve as sensitive experimental probes for the chemical freeze-out conditions at the LHC energy.

\subsection{Sensitivity to the baryon chemical potential} 

Figure~\ref{fig} shows the dependence of different antiparticle/particle ratios, $\bar h/h$, on $\mu_B$ (Fig.~\ref{fig}a) and the dependence of ratios with unequal masses on $T$ (Fig.~\ref{fig}b). For vanishing baryon chemical potential, the density of particles is identical to that of antiparticles, thus $\bar h/h=1$. For finite and increasing baryon chemical potential, the $\bar h/h$ ratios are decreasing functions of $\mu_B$. Such  properties of antiparticle/particle ratios are qualitatively well understood in the Statistical Model. The mass terms in the particle and antiparticle partition functions are identical. Consequently, the leading dependence of the $\bar h/h$ ratios on $\mu_B$  is determined by,
$\bar{h}/h \propto \exp[-2(B \mu_B + S \mu_S)/T], $
where $B$ and $S$ are the  baryon  and strangeness quantum numbers of the particle, respectively. In the  expression above, feed-down contributions from resonance decays  are ignored but they have been included in the model calculations.
In a strangeness neutral system at a fixed temperature and non-vanishing baryon density, the baryon chemical potential is always larger than the strangeness chemical potential. Thus, in the antibaryon/baryon ratios the $\mu_B$-term dominates over the $\mu_S$-term.
This, together with the opposite sign of the quantum numbers $B$ and $S$ of  strange baryons, results in a weaker sensitivity on $\mu_B$ of $\bar{h}/h$ ratios for hadrons with increasing strangeness content. Consequently, as seen in Fig.~\ref{fig}a, at a fixed value of $\mu_B$, the antibaryon/baryon ratios are increasing with the strangeness content of the hadrons. The $\rm \bar{p}/p$ ratio which exhibits the strongest dependence on $\mu_B$ is the ideal observable to  quantify $\mu_B$ experimentally. 
\\
The $\rm K^-/ K^+$ ratio shows qualitatively a similar dependence on the baryon chemical potential. This is due to the strangeness content of the kaons and the fact that $\mu_S$ is an increasing function of $\mu_B$.
For small values of $\mu_B$, in heavy-ion collisions at the LHC energy, the $\bar h/h$ ratios  are expected to be only  weakly dependent on  temperature~\cite{prc}. Thus, these ratios do not  constrain the chemical freeze-out temperature in heavy-ion collisions. 

\subsection{Sensitivity to the temperature at chemical freeze-out and resonance contributions} 

To extract the chemical freeze-out temperature in heavy-ion collisions, one should consider ratios that are composed of hadrons having different  masses. However,  to correctly quantify the temperature dependence of these ratios, one  needs to  include contributions from resonance decays.

Figure~\ref{fig}(c, d) displays particle densities from different sources. Particles called {\sl primary} are directly produced from a thermal system with a given temperature and chemical potentials. The second contribution, called {\sl decay}, originates from short-living resonances which feed the yield of stable hadrons. The sum of both contributions results in the {\sl final} particle density.
\\
There is no feeding to $\Omega^-$ hyperons from heavier resonances. In contrast, only about 50\% of the $\Xi^-$ hyperons are primary, while the other half originates from resonance decays; both contributions exhibit a similar temperature dependence.
For light hadrons like kaons and pions, the feed-down contributions exceed the thermally produced yield. As seen in Fig.~\ref{fig}(c, d) for the thermal conditions at LHC, approximately 75\%  of the pions are expected to be produced by resonance decays. Furthermore, the contributions from resonance decays feature a temperature dependence deviating from that of primarily created hadrons. Therefore, the temperature-driven increase in the heavy/light hadron ratios is progressively diluted when  hadrons with lower masses are considered in the denominator~\cite{prc}.

From the discussions above, it becomes clear that ratios with hyperons, in particular the hyperon/pion ratios, are  excellent observables to extract the chemical decoupling temperature in heavy-ion collisions. Figure~\ref{fig}b shows the temperature  dependence of some particle ratios containing  hyperons.
\\
The largest sensitivity on temperature is exhibited by  the $\Omega^- / \pi^-$ ratio. However,  the  variation is significantly smaller than that expected from only thermally produced particles. This is because at LHC only about 25\% of all pions are expected to be directly produced, while the remaining fraction originates from baryonic and, predominantly, from mesonic resonances~\cite{spencer}. The feed-down contribution to pions increases with temperature and reduces the sensitivity.
\\
In view  of  the   large decay contributions to the  pion yield, the $\Omega^-/$K$^-$ ratio might be a better thermometer of the medium, since the feeding to kaons is noticeably smaller. Figure~\ref{fig}b demonstrates that the two ratios, $\Omega^- / \pi^-$ and $\Omega^-/$K$^-$, are similarly sensitive to temperature variations.
\\
The $\Xi^- / \pi^-$  ratio has  a  similar mass difference as the $\Omega^-/$K$^-$ ratio. However, the very large  feeding of heavy resonances to pions destroys the expected similarity in the temperature dependence of these ratios. The sensitivity of the hyperon to proton ratio on temperature is rather weak due to the small mass difference of constituent particles.

\section{Summary} 
Predictions of the Statistical Model for different particle ratios for Pb-Pb collisions at the LHC energy are presented. The sensitivities of various ratios with respect to the temperature and the baryon chemical potential as well as the contribution of resonances  are discussed. We have shown that  the $\rm \bar{p} / \rm p$ ratio is the best suited observable to extract the value of the baryon chemical potential at chemical freeze-out.  The $\Omega^- / \pi^-$ and the $\Omega^-$/K$^-$ ratio are proposed as thermometers to extract experimentally the chemical freeze-out temperature in central Pb-Pb collisions at LHC.

\end{document}